\documentclass{iaus}
\usepackage{graphicx}

\title[Stellar magnetic cycles] 
{Stellar magnetic cycles}

\author[]   
{A.~F.~Lanza}

\affiliation{INAF-Osservatorio Astrofisico di Catania, Via S. Sofia, 78 -- 95123 Catania, Italy
 \\ email: {\tt nuccio.lanza@oact.inaf.it}}

\pubyear{2010}
\volume{264}  
\pagerange{1--10}
\jname{Solar and Stellar Variability -- Impact on Earth and Planets}
\editors{A.~H.~Andrei, A.~G.~Kosovichev \& J.-P.~Rozelot, Eds.}
\begin{document}

\maketitle

\begin{abstract}
The solar activity cycle is  a manifestation of the hydromagnetic dynamo working inside our star. The detection of activity cycles in solar-like stars and the study of their properties allow us to put the solar dynamo in perspective, investigating how dynamo action depends on stellar parameters and stellar structure. Nevertheless, the lack of spatial resolution and the limited time extension of stellar data pose limitations to our understanding of stellar cycles and the possibility to constrain dynamo models. I  briefly review some results obtained from disc-integrated proxies of stellar magnetic fields and discuss the new opportunities opened by space-borne photometry  made available by MOST, CoRoT, Kepler, and GAIA, and by new ground-based spectroscopic or spectropolarimetric observations. Stellar cycles have a significant impact on the energetic output and circumstellar magnetic fields of late-type active stars which affects the interaction  between  stars and their planets. On the other hand,  a close-in massive planet could  affect the activity of its host star. Recent observations  provide circumstantial evidence of such an interaction with possible consequences  for stellar activity cycles.

\end{abstract}

\firstsection 

\section{Introduction}
The eleven year cycle of  solar activity manifests its effects in the circumsolar and interplanetary environments and has a remarkable impact on  the Sun-Earth relationship. Understanding stellar magnetic cycles will allow us to look at this important phenomenon from a wider perspective. In this review, I shall focus  on a few aspects of solar and stellar cycles that I deem  broadly interesting in this context, although without discussing their impact  on the circumstellar environments and planets. Several reviews have recently covered the topics that I shall not treat here; among others, \cite[Strassmeier (2005)]{Strassmeier05}, \cite[Berdyugina (2005)]{Berdyugina05}, \cite[Hall (2008)]{Hall08}, \cite[Rempel (2008)]{Rempel08}, and \cite[Strassmeier (2009)]{Strassmeier09}. For stellar cycles from X-ray proxies, see, e.g., \cite[G\"udel (2004)]{Gudel04} and \cite[Favata \etal\ (2008)]{Favataetal08}. For the impact of stellar activity on planets see the review by \cite[Ribas (2009)]{Ribas09} in the proceedings of this Symposium.  

\section{The solar cycle and the solar dynamo}
Magnetic fields produce localized perturbations of the flux of energy and momentum in the solar atmosphere that are called {\it active regions} (hereinafter ARs). They are characterized by different phenomena, according to the mean intensity of the magnetic field. Fields of the order of 0.1~T produce cool features in the photosphere, called sunspots, while fields of the order of 0.01~T are associated with a heating of the atmosphere leading to  bright photospheric facular patches and  chromospheric plages. Transient releases of magnetic energy on timescales from a few minutes to hours manifest themselves as solar flares. 
While the formation and evolution of individual active regions are largely unpredictable, their global behaviour reveals a remarkable regularity. The total area  of the sunspots varies in a cyclic manner with a period of $\sim 11$~yr. At the beginning of each cycle, sunspots appear at latitudes of $\approx 35^{\circ}$ in two localized belts, one in each hemisphere, that migrate toward the equator, reaching a latitude of $15^{\circ}-20^{\circ}$ at the  maximum of the total sunspot area and of $5^{\circ}-10^{\circ}$ at the next minimum, when a new couple of active belts appears at high latitudes marking the beginning of a new cycle. In addition to the strong magnetic fields localized inside ARs, the Sun displays a global dipolar magnetic field with an intensity of $\sim 10^{-4}$~T, symmetric around the solar rotation axis, which cyclically reverses its polarities close to the maximum of sunspot activity. \\
\indent
The rotation of the solar convection zone reacts in a cyclic way to the variation of the intensity and spatial distribution of the magnetic field, leading to the phenomenon of the {\it torsional oscillations} (\cite[Howe 2009]{Howe09}). The frequencies of the p-mode oscillations are also affected by the magnetic field and display relative variations of the order of $10^{-4}$ along the 11-yr cycle whose origin is not completely understood yet (cf., e.g., \cite[Salabert \etal\ 2009]{Salabertetal09}). \\
\indent
The time series of sunspot data, which dates back to the XVII century, has revealed that the amplitude and length of the solar cycle are modulated on different timescales, among which the Gleissberg cycle is notable. It consists of a modulation of the amplitude of the 11-yr cycles on a century timescale (e.g., \cite[Kollath \& Olah 2009]{KollathOlah09}).
Investigating time series of proxies extending back to $10^{4}$ yr ago, such as the cosmogenic isotopes  $^{10}$Be or $^{14}$C,  it is possible to investigate the level of solar activity in the distant past. In combination with sunspot data, these studies have revealed intervals of very low solar activity, called {\it grand minima} (see \cite[Usoskin \etal\ 2007]{Usoskinetal07}; \cite[Sokoloff 2009]{Sokoloff09}). At the other end of the time scale, there are short-term cyclic variations of the total sunspot area with periods of $150-160$ days that sometimes appear close to the maximum of the 11-yr cycle (e.g., \cite[Oliver \etal\ 1998]{Oliveretal98}; \cite[Krivova \& Solanki 2002]{krivovaSolanki02}). They were first detected in the occurrence of major flares, which are closely associated to the emergence of new magnetic flux and the formation of new sunspots, by \cite[Rieger \etal\ (1984)]{Riegeretal84} and are therefore called {\it Rieger cycles}. \\
\indent
A theoretical framework to interpret the solar cycle  is offered by dynamo models based on mean-field magnetohydrodynamics (see \cite[Charbonneau 2005]{Charbonneau05}; \cite[Weiss \& Thompson 2009]{WeissThompson09}; and references therein). The global solar dipole field is stretched by the differential rotation giving rise to an intense field in the azimuthal direction which emerges under the action of magnetic buoyancy and leads to the formation of sunspots in the photosphere and magnetic loops in the outer atmosphere. However, the poloidal and  azimuthal fields are subject to a turbulent diffusion, hence the poloidal field will decay, leading eventually to the decay of the azimuthal field, unless some mechanism can regenerate the poloidal field itself. Such a mechanism is much more subtle than differential rotation and is thought to be associated to the action of the Coriolis force on turbulent convection which imparts a non-vanishing mean helicity to the turbulent velocity field of the plasma. Since the magnetic field is frozen into the plasma, such a mean helicity is capable to regenerate a poloidal field from an azimuthal field, thus allowing a dynamo to operate (see, e.g., \cite[Weiss \& Thompson 2009]{WeissThompson09}, and references therein, for details). This is the so-called $\alpha$ effect.  \\ 
\indent
The role of the different effects mentioned above is summarized in the definition of the non-dimensional {\it dynamo number}. Indicating with $\Omega§^{\prime}=\partial \Omega / \partial r$ the typical amplitude of the differential rotation as specified by the radial gradient of the angular velocity (see below), and by $\alpha_{0}$ and $\eta_{\rm t}$ the typical amplitudes of the $\alpha$ effect and the  turbulent diffusivity, respectively, we define the dynamo number as: $D \equiv \alpha_{0} \Omega^{\prime} R^{4}/\eta_{\rm t}^{2}$, where $R$ is the radius of the star. In order for the induction effects represented by the parameters $\alpha_{0}$ and $\Omega^{\prime}$ to overcome the turbulent diffusivity $\eta_{\rm t}$, the dynamo number must be greater than some critical value $D_{\rm c}$, which depends on the boundary conditions, the geometry of the dynamo shell and the spatial distribution of the induction effects. In other words, only when $D \geq D_{\rm c}$ the dynamo can work and we  observe a large-scale magnetic field, otherwise any large-scale field is simply diffused away on a time scale $\tau_{\rm d} \approx R^{2}/\eta_{\rm t}$.
When the magnetic field is amplified enough, it starts to react back on the small scale turbulent motions and the large scale velocity fields. Such a  feedback tends to reduce both the $\alpha$ effect ($\alpha$ quenching) and the differential rotation, thus opposing any further amplification of the field. When a feedback is included, a dynamo model is said to be magnetohydrodynamic or non-linear, in constrast to the simpler linear (or kinematic) models that neglect it and assume a fixed $\alpha$ effect and differential rotation.  \\
\indent
Kinematic dynamo models were proposed long ago for the Sun and account for the oscillation of the sunspot area and the cyclic reversal of the global poloidal field.
They predict that the migration of the azimuthal field as the cycle progresses is in the equatorward direction if the product $\alpha \partial \Omega / \partial r$ is negative in the Northern hemisphere, while it is in the poleward direction if that quantity is positive. At the base of the solar convection zone, helioseismic measurements tell us that 
$\partial \Omega /\partial r > 0 $ at latitudes lower than about $40^{\circ}$, while $\partial \Omega / \partial r < 0$ at higher latitudes (\cite[Thompson \etal\ 2003]{Thompsonetal03}). This  accounts for the observed direction of migration of the activity belts, if a negative $\alpha$ is assumed close to the base of the solar convection zone in the Northern hemisphere (see, e.g. \cite[Schlinchenmaier \& Stix 1995]{SchlinchenmaierStix95}). The scenario is consistent with a localization of the dynamo  in the overshoot layer immediately below the solar convection zone, where a strong shear, mainly in the radial direction, has been detected by helioseismology (e.g., \cite[Thompson \etal\ 2003]{Thompsonetal03}). Another property of such a layer is its subadiabatic stratification that keeps the magnetic field stable until it reaches a critical intensity of  $\approx 1-10$~T, beyond which it emerges toward the photosphere (e.g., \cite[Caligari \etal\ 1995]{Caligarietal95}). Nevertheless, the possibility of a dynamo action distributed throughout the  convection zone or even localized in  the subphotospheric shear layer where $\partial \Omega / \partial r < 0$, is not ruled out. It is possible that the activity observed at the surface of the Sun  results from the contributions of different kinds of dynamos which differ for their location, the storage mechanism of their azimuthal field and the relative amplitude of the field fluctuations with respect to the mean field (cf. \cite[Durney \etal\ 1993]{Durneyetal93}; \cite[Brandenburg 2005]{Brandenburg05}; \cite[Kosovichev 2009]{Kosovichev09}). Finally, it is important to mention the role of the meridional circulation that may act as a sort of conveyor belt transporting the magnetic field between regions where different induction processes take place, e.g., from the overshoot layer, where the shear is mainly localized, into the convection zone, where the $\alpha$ effect is possibly dominant. Therefore, the meridional circulation affects the basic properties of dynamo models including the period of the cycle and the direction and rate of  migration of the mean azimuthal field. \\
\indent
A fundamental question concerns the possibility of understanding a complex and highly non-linear system such as the solar (or a stellar) dynamo by means of simple parametric models like mean-field models. They drastically reduce the number of degrees of freedom of the system which are  of the order of $10^{27}$ (\cite[Canuto 2000]{Canuto00}) and apply highly simplified parametric dependences to account for non-linear effects. Yet, they seem to be quite successful in explaining the main characteristics of the solar cycle. The reason for their success and the degree up to which we may hope to extend the comparison between models and observations has been recently discussed by Spiegel (2009).  

\section{Stellar activity cycles}
\subsection{Stellar cycles from chromospheric proxies}
Stellar discs are not resolved by current observations, therefore, we need  suitable proxies for stellar magnetic fields that can be derived from disc-integrated photometric or spectroscopic measurements. In the Sun, the transits of ARs across the disc during its rotation produce characteristic flux variations. 
Cool spots produce  flux decreases while faculae induce an increase of the flux with a remarkable centre-to-limb dependence because their contrast is greater close to the limb. The Total Solar Irradiance, i.e., a measure of the solar bolometric flux at the distance of 1 AU, is a good proxy of the optical flux modulation induced by ARs. Its variations  have now been monitored for almost three  activity cycles. Owing to the predominant effect of the faculae, the Sun at the maximum of activity is brighter by $ \sim 0.1$ percent than at the minimum (\cite[Fr\"ohlich \& Lean 2004]{FrohlichLean04}). In the case of solar-like stars, it has been difficult to measure variations of the order of 0.001 mag from the ground over timescales of decades. Therefore, the use of long-term series of optical photometry to study  activity cycles in  stars with an activity level comparable to that of the Sun is still quite limited. The situation is different in the case of  chromospheric proxies, notably the excess flux in the core of the Ca~II~H\&K lines, which varies by $\approx 10$ percent along the solar cycle (\cite[Hall \& Lockwood 2004]{HallLockwood04}). Such a variation is  measured by means of the so-called $S$ index (\cite[Vaughan \etal\ 1978]{Vaughanetal78}), whose systematic monitoring in a sample of $\sim 100$ late-type  main-sequence single stars started $\sim 40$ years ago. This is the famous  H\&K project dedicated to the detection and study of stellar activity cycles. Its main results have been summarized by \cite[Baliunas \etal\ (1995, 1998)]{Baliunasetal95,Baliunasetal98}. They find that about 60 percent of  main-sequence stars with spectral types from F to early M show cycles. Cycles shorter than $\sim 7$ yr are never observer among stars with a rotation period longer than $\approx 15-20$ days. Several stars show two independent periodicities. 
About 25 percent of the sample show a variable $S$ index, but without any obvious periodicity. This behaviour is more frequent among  younger and more rapidly rotating stars, while stars with an age comparable to the Sun tend to show  a cyclic behaviour or an almost constant flux. The latter have been dubbed {\it flat activity stars} and amount to $\sim 15$ percent of the sample. It was suggested that they could be in a phase of extended activity minimum, akin the Sun during its grand minima. However,  recent studies have questioned such an interpretation, also because they span a range of chromospheric flux from below the solar minimum to near the higher levels of the present solar cycle (\cite[Hall \& Lockwood 2004]{HallLockwood04}). Moreover, the $S$ index is not a good proxy of magnetic activity at low levels of magnetic flux because the core fluxes of the Ca~II~H\&K lines are dominated by the basal chromospheric heating due to acoustic waves produced by photospheric convection, irrespective of the magnetic flux. It is only when the chromospheric flux is significantly higher than the basal level for a given spectral type (and luminosity class) that the Ca~II~H\&K lines are useful to study magnetic activity and cycles. Such considerations are important also for subgiant and giant stars which show a smaller frequency of cyclic variations ($\approx 40$ percent) and a more variable level of  activity (\cite[Baliunas \etal\ 1998]{Baliunasetal98}), possibly because their chromospheric flux is dominated by acoustic heating. \\
\indent
Chromospheric proxies can be used  to measure the rotation period of the ARs in a star, when they have a sufficiently long lifetime to produce a unambiguous  rotational modulation of the flux. Therefore, it is possible to correlate the level of chromospheric activity with the rotation period. Observations of the Sun as a star showed a variation of the rotation period in phase with the solar cycle, i.e., a decrease of the rotation period as the cycle progressed from minimum to maximum and then a further reduction followed by an abrupt increase when the activity went through the next minimum and a new cycle began (\cite[Donahue \& Keil 1995]{DonahueKeil95}). This was  consistent with a migration of the activity belts from the higher latitudes toward the equator, followed by the appearance of new high latitude belts at the beginning of a new cycle. Nevertheless, the solar-type correlation is not the only one observed among the H\&K project stars. It has been quite surprising to find that several stars show an {\it anti-solar} behaviour, i.e., the rotation period of their ARs increases as the cycle progresses  from minimum to maximum activity (see, e.g., the case of HD~114710; \cite[Donahue \& Baliunas 1992]{DonahueBaliunas92}). It is unlikely that such stars possess an anti-solar pattern of surface differential rotation, i.e., with the poles rotating faster than the equator (e.g., \cite[R\"udiger \etal\ 1998]{Rudigeretal98}). Therefore, a possible explanation is that their ARs migrate poleward instead than equatorward. This may  happen if stellar activity is not confined to latitudes close to the equator, but is well extended toward the poles. If the dynamo operates in a stellar overshoot region at high latitudes, the direction of migration of the dynamo wave is expected to be poleward in the case of an internal rotation profile similar to that of the Sun because $\partial \Omega / \partial r <0$ at latitudes higher than $\sim 40^{\circ}$ (\cite[Thompson  \etal\ 2003]{Thompsonetal03}). Of course, there can be another dynamo wave propagating  at low latitudes toward the equator, but the poleward wave may dominate the observed stellar activity if the inclination of the stellar rotation axis is low, i.e., the star is viewed approximately pole-on. \\
\indent
As a development of such studies, \cite[Baliunas \etal\ (2006)]{Baliunaset06}  proposed  an interpretation for the different shapes of the modulation of the $S$ index vs. time in stars showing cycles. 
Stars with a low level of activity often show an almost sinusoidal modulation, while more active objects display anharmonic variations. \cite[Baliunas \etal\ (2006)]{Baliunaset06} attribute such a difference to the different dynamo regimes occurring in different stars. Objects with a dynamo number close to the critical value  
$D \sim D_{\rm c}$ are characterized by an almost sinusoidal variation of the $S$ index and by a migration of the activity belts in latitude. On the other hand, stars with 
$D \gg D_{\rm c}$ show cycles with a markedly non-sinusoidal modulation and their magnetic field does not migrate  appreciably in latitude as their cycle progresses, i.e., there is a weak correlation between the rotation period of their ARs and the phase of the cycle. This interpretation is based on a  dynamo model including an $\alpha$-quenching mechanism. It assumes that in stars with a high level of activity the $\alpha$ effect is negligible for most of the cycle because the magnetic field intensity is so high as to  strongly quench the mean helicity of the turbulence. In this regime, the dynamo wave does not migrate appreciably because its rate and direction of migration are controlled by the quantity $\alpha \partial \Omega / \partial r$ which is close to zero. Only when the turbulent diffusion of the poloidal field eventually affects the strength of the azimuthal field, the $\alpha$ effect is resumed to allow a regeneration of the former. However, this occurs only for a limited interval of time during each cycle and has no  consequences for the migration of the dynamo wave which stays almost stationary. The  prolonged suppression of the $\alpha$ effect leads to a remarkably anharmonic variation of the mean magnetic field which implies an anharmonic variation of the $S$ index.  \\
\indent
Finally, interesting results have been derived by a comparison of the chromospheric long-term variations with the seasonal optical magnitudes of several H\&K project stars for which   time series of $10-20$ yr are now available  (\cite[Lockwood \etal\ 2007]{łockwwodetal07}; \cite[Hall \etal\ 2009]{Halletal09}). They show two different kinds of behaviour: a) there are stars similar to the Sun which become brighter in the optical (and probably in the bolometric passband) when they are more chromospherically active; and, b) there are stars which become fainter when they are more chromospherically active. The former have a chromospheric activity level comparable to that of the Sun, so their long-term optical variations are dominated by  faculae.  The latter are more active than the Sun and their ARs are  dominated by  cool spots, so they get fainter when their activity increases. 
\subsection{Stellar cycles from photospheric proxies}
The variability of the optical flux can be easily monitored from the ground to search for activity cycles  if its amplitude is greater than a few 0.01 mag. In this way, \cite[Messina \& Guinan (2002)]{MessinaGuina02}  studied cycles in a sample of young solar analogues with spectral types G0V-G5V and ages ranging from $\approx 130$ to $\approx 700$ Myr. Although the sample was limited, it was very interesting to study the early evolution of solar activity. They found that the fastest rotating stars  tended to display longer cycles, with the range of the observed cycle periods apparently converging from a large dispersion at ages of 
$\approx 100-200$ Myr toward the solar cycle period at the age of the Sun.  \cite[Messina \& Guinan (2003)]{MessinaGuinan03} investigated the correlation between the period of the rotational modulation  and the phase of the  cycle in the above sample of stars finding a clearly significant correlation in two of them, and a less significant correlation in the other four. Both solar-like and anti-solar behaviours were observed, as in the case of the chromospheric cycles of the H\&K project sample. 
The amplitude of the optical variability on timescales from the rotation period to the cycle period, was well correlated to the inverse of the \textit{Rossby number} of each star. The Rossby number is defined as: $Ro \equiv P_{\rm rot}/\tau_{\rm c}$, where $P_{\rm rot}$ is the rotation period of the star and $\tau_{\rm c}$ the correlation time of the turbulent motions at the base of the stellar convection zone, usually set equal to the convective turnover time one mixing length above the base of the convection zone, as computed by means of the standard  mixing-length theory.  Assuming that $\alpha_{0} = \Omega l$,  $\Omega^{\prime} \propto \Omega$, and
$\eta_{\rm t} = l^{2} / \tau_{c}$, where $l$ is the mixing length at the base of the stellar convective envelope, we find $D \propto Ro^{-2}$.
Therefore, the Rossby number is a measure of the dynamo number which can be derived  by combining observations of the stellar rotational modulation with a  model of its internal structure (see, e.g., \cite[Messina \& Guinan 2003]{MessinaGuinan03}). \\
\indent
Late-type stars in close binary systems are forced into fast rotation by  tidal interactions and show a very high level of magnetic activity. Among those systems, the most active are  the RS Canum Venaticorum binaries. They are detached systems consisting of a G or K subgiant, and a dwarf or subgiant of spectral type F or G. Their phenomenology has been described in several reviews (e.g., \cite[Rodon\`o 1992]{Rodono92}; \cite[Guinan \& Gimenez 1993]{GuinanGimenez93}).  Long-term optical monitoring has revealed cycles of $10-20$ yr in several systems (e.g., \cite[Rodon\`o \etal\ 1995]{Rodonoetal95}; \cite[Henry \etal\ 1995]{Henryetal95}; \cite[Ol\'ah \etal\ 1997]{Olahetal97}; \cite[Lanza \etal\ 1998]{Lanzaetal98}; \cite[Rodon\`o \etal\ 2000]{Rodonoetal00}; \cite[Lanza \etal\  2001]{Lanzaetal01}; \cite[Lanza \etal\ 2002]{Lanzaetal02}; \cite[Lanza \etal\ 2006]{Lanzaetal06}; \cite[Berdyugina \& Henry 2007]{”erdyuginaHenry07}). Recently, \cite[Buccino \& Mauas (2009)]{BuccinoMauas09} have investigated chromospheric cycles in some RS~CVn systems using long-term UV data from the IUE satellite; and \cite[Messina (2008)]{Messina08} has discussed the long-term variability of the optical colours of several systems finding in some cases evidence of a predominance of the faculae at the maximum of activity. The long-term optical variability of RS~CVn systems is generally  anharmonic, in agreement with the dynamo models by \cite[Baliunas \etal\ (2006)]{Baliunaset06} for very active stars. Moreover,  changing of the cycle period as well as multiperiodic variations are often observed (cf.  \cite[Ol\'ah \etal\ 2009]{Olahetal09}), which may be related to the complex feedback mechanisms operating in a highly non-linear stellar dynamo (cf. \cite[Spiegel 2009]{Spiegel09}). \\
\indent
The binary nature of RS CVn stars allows us to look for other proxies of magnetic field variations. A very interesting candidate is the long-term variability of their orbital period which shows cycles with relative amplitudes of the order of $10^{-5}$ on time scales from a few decades up to a few centuries. The same phenomenon is observed in semi-detached binaries, such as Algols and  cataclysmic variables. The latter are particularly interesting because they show the phenomenon also when their secondary components are fully convective (\cite[Ak \etal\ 2001]{Aketal01}; \cite[Baptista \etal\ 2008]{Baptistaetal08}), thus providing a  circumstantial evidence of magnetic cycles in fully convective dwarf stars with a sufficiently fast rotation. For more information on the connection between orbital period modulation and dynamo action see, e.g., \cite[Lanza \etal\ (1998)]{Lanzaetal98}, \cite[Lanza \& Rodon\`o (1999)]{LanzaRodono99}, \cite[Lanza \& Rodon\`o (2004)]{LanzaRodono04}, \cite[Lanza (2005)]{Lanza05}, and \cite[Lanza (2006a)]{Lanza06a}. 
\subsection{Dependence of the cycle period on stellar parameters}
Since stellar cycles are  manifestations of  stellar dynamos whose regimes mainly depend on the dynamo number $D$, most of the studies have focussed on the relationship between the period of the cycle and the rotation period, or the Rossby number, assumed to be a better proxy for $D$. A plot of the cycle period $P_{\rm cyc}$ vs. the rotation period $P_{\rm rot}$ shows a large scatter, particularly for $P_{\rm rot} < 10-20 $ days and a generally weak correlation between $P_{\rm cyc}$ and $P_{\rm rot} $ (cf., e.g., \cite[Saar \& Brandeburg 1999]{SaarBrandenburg99}; \cite[Ol\'ah \etal\ 2009]{Olahetal09}). When the cycle period is measured in units of the rotation period of the star and plotted vs. the Rossby number, the correlation improves if the stars are subdivided into three branches (cf. Fig. 5 of \cite[Saar \& Brandenburg 1999]{SaarBrandenburg99}). Those with a low or intermediate level of activity, i.e., pratically all the targets of the H\&K project, occupy two roughly parallel branches in the plot and show a $P_{\rm cyc}/P_{\rm rot}$  decreasing with the increase of the inverse Rossby number. On the other hand, very active stars, such as those in RS CVn binaries, show an increase of $P_{\rm cyc}/P_{\rm rot}$ with the increase of the inverse Rossby number. Their branch is connected to that of the intermediate active stars by several objects that may be regarded as being in a transitional evolutionary stage from very high activity to intermediate activity. The six young solar analogues investigated by \cite[Messina \& Guinan (2002)]{MessinaGuinan02} fall on the branches corresponding to intermediate or highly active stars, with one object in the transitional stage. \\
\indent
This complex scenario is still not satisfactorily understood. Although some conjectures have been proposed to account for the different branches, such as  changes in the $\alpha$ effect with the field strength (cf. \cite[Saar \& Brandenburg 1999]{SaarBrandenburg99}), or  different locations and aspect ratios of the dynamo shell (e.g., \cite[Bohm-Vitense 2007]{BohmVitense07}), it is difficult to constrain the dynamo models. 

\section{New perspectives opened by the latest and forthcoming observations}
The recent launch of the satellite CoRoT has opened a new perspective for the study of stellar activity. The MOST  satellite had already shown the potentiality of space-borne photometers in terms of precision, stability and high-duty cycle (cf., e.g., \cite[Walker \etal\ 2007; 2008]{Walkeretal07,Walkeretal08}), but CoRoT with its extended time series  allowed us to find  more interesting results, notably in the case of active stars accompanied by hot Jupiters.  Spot modelling techniques based on high-precision photometry have been validated through an extensive application to solar irradiance time series (cf. \cite[Lanza \etal\ 2007]{Lanzaetal07}). \\
\indent
The young active G7V star hosting CoRoT-2b has  $P_{\rm rot} \sim 4.52$ days and a spot modelling of its light curve has revealed a short-term spot cycle of $\sim 29$ days, akin the solar Rieger cycles (\cite[Lanza \etal\ 2009a]{Lanzaetal09a}). The shorter period may be a consequence of the faster stellar rotation in the framework of the model by \cite[Lou (2000)]{Lou00} that interprets Rieger cycles as a manifestation of Rossby-type waves trapped in the outer layers of the convection zone. An alternative explanation calls into play a possible influence of the close-in hot Jupiter because the cycle period is very close to ten synodic periods of the planet with respect to the mean stellar rotation period.  
Indeed, MOST and CoRoT are  providing hints of a possible influence of close-in giant planets on stellar magnetic activity, revealing persistent  ARs that rotate synchronously with the planet in  systems such as $\tau$ Bootis (\cite[Walker \etal\ 2008]{Walketetal08}) and CoRoT-4 (\cite[Lanza \etal\ 2009b]{Lanzaetal09b}). More on these intriguing phenomena
can be found in \cite[Lanza (2008)]{Lanza08}, \cite[Shkolnik \etal\ (2009)]{Shkolniketal09}, and \cite[Lanza (2009)]{Lanza09}.   \\
\indent
CoRoT time series will allow us several more detections of Rieger-like cycles and possible interactions between stars and their planets because more than 100,000 late-type stars will be monitored by the end of the mission for time intervals ranging from 20 to 150 days. To date, apart from the Sun and CoRoT-2a, only the active close binary  UX Arietis has been suggested to show  Rieger cycles (\cite[Massi \etal\  2005]{Massietal09}). On the other hand, the recently launched Kepler satellite will allow us to extend continuous photometric monitoring to  time intervals of 3-4 yr, improving the photometric precision by at least a factor of 3. It should be capable of monitoring the activity cycle variations of p-mode eigenfrequencies in stars brighter than $V \sim 9$, hopefully providing us with information for a better understanding of the phenomenon (e.g., \cite[Karoff \etal\ 2009]{Karoffetal09}).  Another interesting perspective shall be opened by GAIA which will observe the field of Kepler for $2-4$ years after the end of the Kepler mission, although with a very sparse time sampling (on the average 90 points in five years), allowing us to get some information on decade-long magnetic cycles in stars brighter than $V \approx 15$. \\ 
\indent
Rich opportunities are opened by the new ground-based spectroscopic and spectropolarimetric observations. A database of Doppler Imaging maps of very active  stars is now available, in some cases with an extension of $10-15$ years,  allowing us an interesting comparison with long-term photometry (see, e.g., \cite[Strassmeier 2009]{Strassmeier09} for a  discussion of Doppler Imaging results). Among the many studies based on Doppler Imaging, I would like to mention the detection of long-term oscillations of the amplitude of surface differential rotation in the very young ($\approx 30$ Myr) K0 dwarf AB Dor (\cite[Collier Cameron 2007]{CollierCameron07}). Such studies allow us to investigate the back reaction of the Lorentz force on stellar rotation in highly non-linear dynamo regimes and can be analyzed on the basis of general models for the angular momentum transport inside a convection zone to derive quantitative estimates of the internal torques and magnetic fields (see \cite[Lanza 2006b]{Lanza06b}, \cite[Lanza 2007]{Lanza07}). \\
\indent
Spectropolarimetric observations of $\tau$ Bootis have revealed two consecutive reversals of its photospheric magnetic field, suggesting a solar-like magnetic cycle with a period of only $\sim 2$ yr, i.e., ten times shorter than the solar magnetic cycle (\cite[Fares \etal\  2009]{Faresetal09}). Such a very short cycle could be  related to the influence of a close-in giant planet orbiting $\tau$ Boo, as conjectured by \cite[Lanza (2008; 2009)]{Lanza08,Lanza09}. So far $\tau$ Boo is the only star  to have shown a complete cycle of polarity reversals, apart from the Sun. A long-term monitoring of the highly active K1IV component of the  RS CVn system HR~1099 did not show such a phenomenon, although the observations cover almost 15 years (\cite[Donati \etal\ 2003]{Donatietal03}; \cite[Petit \etal\  2004]{Petitetal04}). 
\section{Conclusions}
A major limitation to our understanding of stellar magnetic cycles is  the lack of spatial resolution of stellar discs and asteroseismic measurements of internal rotation. Therefore, our picture is not as detailed as in the case of the Sun. In spite of that, the detection and study of stellar cycles by proxies is now a well established field of research and long-term time series have revealed the existence of a complex scenario. Stellar cycles can be associated with quasi-sinusoidal or highly anharmonic variations of the proxies, they show single or  multiple periodicities,  and their periods may change versus time (cf. \cite[Ol\'ah \etal\ 2009]{Olahetal09}). The relationship of the cycle period to the stellar parameters is by no means a simple one. Even considering the correlation with the Rossby number, which provides us with a measure of the dynamo number, no simple scenario emerges. Present dynamo models have difficulties in accounting for such a complex correlation, so they are not significantly constrained by it. A significant progress may come from the new  long-term spectroscopic and spectropolarimetric observations that can measure surface differential rotation and photospheric magnetic fields in highly active stars, i.e., young main-sequence stars or late-type components of close active binaries. Such observations provide a direct view of the large scale topology of the magnetic field in the photosphere and can be used to estimate the intensity of the internal field by measuring its back-reaction on stellar rotation. For slowly rotating, sun-like stars a new perspective is opened by  CoRoT and Kepler high-precision photometric time series. They have already allowed us to detect  Rieger-like cycles in the planet-hosting star CoRoT-2a and have provided hints of a possible influence of close-in giant planets on the activity of their host stars. 


\end{document}